\documentclass[aps,prl,twocolumn,reprint,
			   groupedaddress,superscriptaddress,
			   amsfonts,amssymb,amsmath, 
			   citeautoscript,
			   a4paper
			   ]{revtex4-1}

\usepackage{microtype} 

\usepackage{txfonts}  
\usepackage{txfontsb} 
\usepackage{bm} 

\usepackage{xcolor}
\usepackage{graphicx}

\usepackage{hyperref}
\hypersetup{
    colorlinks,
    linkcolor={blue!75!black!80!yellow},
    citecolor={blue!75!black!80!yellow},
    urlcolor={blue!75!black!80!yellow}
}

\usepackage[centering,hmargin=2cm,vmargin=2cm,tmargin=34mm,bmargin=42mm]{geometry}

\color{black!88!white}

\newcommand{\rv}{\mathbf{r}}
\newcommand{\ef}{\epsilon_{\text{\textsc{f}}}}
\newcommand{\vf}{v_{\text{\textsc{f}}}}
\newcommand{\tsc}[1]{\text{\textsc{#1}}}
\newcommand{\e}{\mathrm{e}}

\newcommand{\appropto}{\mathrel{\vcenter{
  \offinterlineskip\halign{\hfil$##$\cr
    \propto\cr\noalign{\kern.2pt}\sim\cr\noalign{\kern-2.5pt}}}}}

\begin{document}
\title{\color{black!88!white} Classical and Quantum Plasmonics in Graphene Nanodisks: the Role of Edge States}

\author{Thomas~Christensen}
\affiliation{Department of Photonics Engineering, Technical University of Denmark, DK-2800 Kgs. Lyngby, Denmark}
\affiliation{Center for Nanostructured Graphene, Technical University of Denmark, DK-2800 Kgs. Lyngby, Denmark}
\author{Weihua~Wang}
\affiliation{Department of Photonics Engineering, Technical University of Denmark, DK-2800 Kgs. Lyngby, Denmark}
\affiliation{Center for Nanostructured Graphene, Technical University of Denmark, DK-2800 Kgs. Lyngby, Denmark}
\author{Antti-Pekka~Jauho}
\affiliation{Center for Nanostructured Graphene, Technical University of Denmark, DK-2800 Kgs. Lyngby, Denmark}
\affiliation{Department of Micro- and Nanotechnology, Technical University of Denmark, DK-2800 Kgs. Lyngby, Denmark}
\author{Martijn~Wubs}
\affiliation{Department of Photonics Engineering, Technical University of Denmark, DK-2800 Kgs. Lyngby, Denmark}
\affiliation{Center for Nanostructured Graphene, Technical University of Denmark, DK-2800 Kgs. Lyngby, Denmark}
\author{N.~Asger~Mortensen}
\email{asger@mailaps.com}
\affiliation{Department of Photonics Engineering, Technical University of Denmark, DK-2800 Kgs. Lyngby, Denmark}
\affiliation{Center for Nanostructured Graphene, Technical University of Denmark, DK-2800 Kgs. Lyngby, Denmark}


\begin{abstract}\color{black!88!white}

Edge states are ubiquitous for many condensed matter systems with multicomponent wave functions. For example, edge states play a crucial role in transport in zigzag graphene nanoribbons.  Here, we report microscopic calculations of quantum plasmonics in doped graphene nanodisks with zigzag edges. 
We express the nanodisk conductivity $\sigma(\omega)$ as a sum of the conventional bulk conductivity $\sigma_{\tsc{b}}(\omega)$, and a novel term $\sigma_{\tsc{e}}(\omega)$, corresponding to a coupling between the edge and bulk states. We show that the edge states give rise to a red-shift and broadening of the plasmon resonance, and that they often significantly impact the absorption efficiency.  We further develop simplified models, incorporating nonlocal response within a hydrodynamical approach, which allow a semiquantitative description of plasmonics in the ultrasmall size regime. However, the polarization dependence is only given by fully microscopic models. The approach developed here should have many applications in other systems supporting edge states.

\end{abstract}

\maketitle

\color{black!88!white}

\emph{Introduction.---}Plasmonics at the nanoscale introduces a host of novel phenomena, both in terms of improved efficacy of certain classical phenomena, e.g.\ extreme field enhancements, but also conceptually by offering a tunable transition from the classical to the quantum regime~\cite{Tame:2013}. Probing and understanding this transition in detail, and in particular the breakdown of classical predictions, is an important task in view of the progress in nanofabrication~\cite{Ostrikov:2011,Henzie:2009}. With the emergence of low-dimensional materials such as graphene, new avenues develop, both experimentally and theoretically. Graphene, and several other low-dimensional systems, exhibits an approximately linear, gapless, two-band energy dispersion $\epsilon = \pm\hbar\vf k$, with Fermi velocity $\vf$. The plasmonic consequences of this nonstandard dispersion and dimensionality have been investigated vigorously in recent years~\cite{Grigorenko:2012,Abajo:2014,Jablan:2009,Koppens:2011,Brar:2013,YanLow:2013}.

The accurate effective description of low-energy excitations in graphene by simple tight-binding (TB) Hamiltonians allows investigations of nonclassical plasmonic features of relatively large graphene structures~\cite{Thongrattanasiri:2012,Thongrattanasiri:2013}. It has recently been theoretically demonstrated that the optical excitations of few-atom graphene nanostructures involve multiple individual electron-hole pairs (EHPs) strongly modified by the Coulomb interaction, occasionally referred to as molecular plasmons~\cite{Manjavacas:2013}. Conversely, experimental measurements on ensembles of larger disks, of radii $R\gtrsim  50\ \mathrm{nm}$, have exhibited distinctly classical features~\cite{Fang:2012,Zhu:2014}.

\begin{figure}
\centering
\includegraphics[scale=.62]{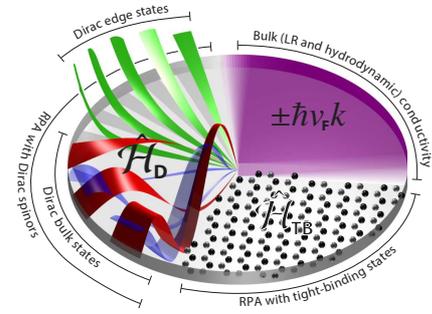} 
\caption{Illustration of considered levels of approximation for a graphene nanodisk. Angular slices of Dirac ZZ bulk state spinor-components are indicated in red and blue, and edge state nonzero-components in green.\label{fig:methods}}
\end{figure}

In this paper we show that for smaller graphene disks, though larger than $R \gtrsim 7\ \mathrm{nm}$, two essential modifications of the classical single-disk response arise, due to edge states and to nonlocal response, producing an overall redshift and broadening of the dipole resonance.
In particular, we show that the existence of edge states due to zigzag (ZZ) features can be accounted for via an edge-state conductivity, whilst the effect of nonlocal response can be accounted for effectively within a hydrodynamic model. This affirms and extends the supposition regarding the crucial role of edge states in prior numerical work~\cite{Thongrattanasiri:2012}. In Fig.~\ref{fig:methods} we outline and summarize the different computational approaches considered in this paper.

\emph{Electronic states.---}The simplest atomistic description of the conduction electrons of graphene, without explicit treatment of spin, is given by the p$_{\text{z}}$-orbital nearest-neighbor TB Hamiltonian with hopping energy $t_{\text{\textsc{ab}}}=2.8\ \mathrm{eV}$:
\begin{equation}\label{eq:tbhamiltonian}
\hat{\mathcal{H}}_{\text{\textsc{tb}}} = -t_{\text{\textsc{ab}}}\sum_{\langle j,j'\rangle}\!\!\hat{a}_j^\dagger \hat{b}_{j'}+\hat{b}_{j'}^\dagger \hat{a}_j,
\end{equation}
with A- and B-sublattice annihilation (creation) operators $\hat{a}_j^{(\dagger)}$ and $\hat{b}_j^{(\dagger)}$ at sites $j$, and with $\langle j,j'\rangle$ indicating summation over nearest neighbors. In the low-energy limit, for extended graphene, the characteristics of the TB approach are asymptotically reproduced by the four-spinor Dirac equation, $\hat{\mathcal{H}}_{\text{\textsc{d}}}\bm{\psi}(\rv) = \epsilon\bm{\psi}(\rv)$, with the Hamiltonian~\cite{Neto:2009}
\begin{equation}\label{eq:dirachamiltonian}
\hat{\mathcal{H}}_{\text{\textsc{d}}}  = \vf (\tau_0\otimes\sigma_x \hat{p}_x + \tau_z\otimes\sigma_y \hat{p}_y),
\end{equation}
where $\bm{p}=-i\hbar\nabla$ denotes momentum, and with Pauli matrices $\tau_i$ and $\sigma_i$ belonging to valley- and sublattice-subspaces, respectively. In the absence of valley-mixing, the four-spinor equation for
$\bm{\psi}(\rv) = [\psi_{\tsc{a}}^{\bm{K}},\psi_{\tsc{b}}^{\bm{K}},\psi_{\tsc{a}}^{\bm{K}'},\psi_{\tsc{b}}^{\bm{K}'}]^{\tsc{t}}$ decouples into a pair of two-spinor equations for $\bm{\psi}^{\kappa}(\rv) = [\psi_{\tsc{a}}^{\kappa}(\rv),\psi_{\tsc{b}}^{\kappa}(\rv)]^{\tsc{t}}$ associated with Dirac valleys $\bm{K} = [1,1/\!\sqrt{3}]^{\tsc{t}}2\pi/3a$ for $\kappa = 1$ and $\bm{K}' = [1,-1/\!\sqrt{3}]^{\tsc{t}}2\pi/3a$ for $\kappa = -1$~\cite{Note1}.

Finite graphene structures are easily modeled with Eq.~\eqref{eq:tbhamiltonian} by omitting the absent neighbors in the matrix representation of $\hat{\mathcal{H}}_{\tsc{tb}}$, whose dimension equals the number of constituent carbon atoms.
For the continuum Dirac equation, Eq.~\eqref{eq:dirachamiltonian}, suitable boundary conditions (BCs) are needed. General considerations, enforcing no-spill current conditions, hermiticity, and unitarity, lead to a rather broad family of allowable BCs~\cite{McCann:2004,Akhmerov:2008a}, which can be made explicit by using the atomistic details of the structural termination. In the present work we consider ZZ lattice termination (which can be considered appropriate in general for non-armchair minimal lattice terminations as argued in Ref.~\onlinecite{Akhmerov:2008a}) forcing a single sublattice component to vanish, e.g.\ forcing $\psi_{\tsc{a}}^{\kappa} (\rv)= 0$ on the boundary if the ZZ edge belongs to the B-sublattice.
For comparison we also consider the infinite mass (IM) BC~\cite{Berry:1987}, corresponding microscopically to confinement due to an atomically staggered potential~\cite{Akhmerov:2008a}, which enforces an intersublattice phase-relationship $\psi_{\tsc{b}}^{\kappa}(\rv)/\psi_{\tsc{a}}^{\kappa}(\rv) = i\e^{i\kappa\theta}$, with $\theta$ denoting the tangential boundary angle~\cite{Peres:2009}.

\begin{figure}[htb!]
\hspace*{-.4em}\includegraphics[scale=.93]{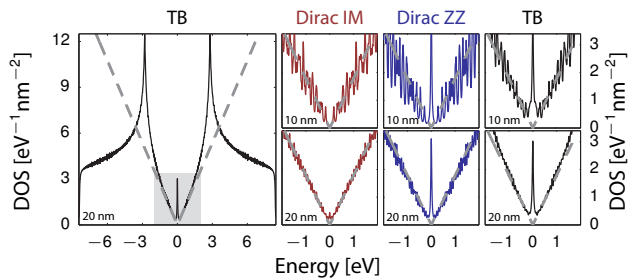} 
\caption{Density of states for graphene nanodisk in Dirac ZZ and IM, and TB treatments, broadened by a loss of $\hbar\eta = 24\ \mathrm{meV}$, with disk-diameter indicated. The asymptotic low-energy, bulk graphene DOS is indicated in dashed gray. The region of approximately linear DOS is indicated by gray shading.\label{fig:dos}}
\end{figure}

Upon application of BCs, the otherwise linear Dirac dispersion $\epsilon = \hbar\vf k$ is transformed into a discrete set of energies and associated spinors. For the case of a homogeneous disk of radius $R$, the nonzero-energy spinors are quantized in angular and radial quantum numbers $l=0,\raisebox{.8pt}{$\pm$} 1, \raisebox{.8pt}{$\pm$} 2, \ldots$ and $n = 1,2,\ldots$~\cite{Wunsch:2008,Grujic:2011a}:
\begin{equation}\label{eq:bulkstates}
\bm{\psi}_{ln}^{\kappa}(\tilde{r},\theta) = \frac{\e^{i l\theta}}{\sqrt{N_{ln}^{\kappa}}}
\begin{bmatrix}
J_l(\beta_{ln}\tilde{r})\\
i\kappa J_{l+\kappa}(\beta_{ln}\tilde{r})\e^{i\kappa\theta}
\end{bmatrix},
\end{equation}
with normalization $N^{\kappa}_{ln}$, see Supplemental Material (SM)~\cite{Note2}, expressed through the dimensionless radial coordinate $\tilde{r} = r/R$ and momenta $\beta_{ln} = \epsilon_{ln}/{\hbar\omega_R}$ with circumferential fermion frequency $\omega_R = \vf/R$. The ZZ BC energies are valley-independent and correspond to zeros of the Bessel function, i.e.\ $\beta_{ln}$ fulfills $J_l(\beta_{ln}) = 0$, while the IM BC yields valley-dependent energies, given by $\kappa J_{l+\kappa}(\beta_{ln}^{\kappa}) = J_l(\beta_{ln}^\kappa)$. Additionally, for the ZZBC a set of zero-energy spinors, here denoted by $\bm{\phi}^{\kappa}$, exist with angular quantum numbers $\ell = 0,1,\ldots,\ell_{\mathrm{max}}$~\cite{Wunsch:2008,Grujic:2011a}:
\begin{equation}\label{eq:edgestates}
\bm{\phi}^{\kappa}_{\ell}(\tilde{r},\theta) = \frac{\e^{-i\kappa \ell\theta}}{\sqrt{\mathcal{N}_{\ell}}}
\begin{bmatrix}
0\\ \tilde{r}^{\ell}
\end{bmatrix},
\end{equation}
with normalization $\mathcal{N}_{\ell}$, see SM. The phenomenologically introduced cutoff angular quantum number $\ell_{\mathrm{max}}$ is required to avoid a divergence of the density of states at zero energy, and is chosen to ensure a total number of zero-energy states (including spin and valley-degeneracy) $\sim 2\pi R/3a_{\tsc{lc}}$~\cite{Akhmerov:2008a}, with lattice constant $a_{\tsc{lc}}=2.46\ \AA$, see SM.
In Fig.~\ref{fig:dos} we show the resulting non-interacting density of states, phenomenologically broadened by an electron collision rate $\eta$, computed as $\mathrm{DOS}(\epsilon) = \tfrac{2}{\pi\mathcal{A}}\sum_{\nu} \mathrm{Im}\big[(\epsilon_{\nu}-\epsilon-i\hbar\eta)^{-1}\big]$ with $\mathcal{A}$ denoting sample area and with $\sum_\nu$ denoting summation over all states $\nu$ (excluding spin, which contributes a factor 2).  Also, the DOS for a TB-model is shown, for a bond-centered disk. 
A key feature of both TB and Dirac ZZ treatments is a prominent peak at zero-energy associated with edge states, which is not reproduced in either Dirac IM or in bulk approximations. Additionally, due to breaking of valley and azimuthal symmetry in TB the inter-state energy-level spacing is overestimated in Dirac treatments relative to TB. Nevertheless, the total number of edge and bulk states in Dirac ZZ and TB is in good agreement, see SM. Due to the absence of edge states in Dirac IM vis-\`{a}-vis its presence in TB, we focus in the following on Dirac ZZ. Finally, we note the complete absence of non-conical dispersion effects, e.g.\ trigonal warping, in the Dirac treatment, whose exemption, however, is expected to be unimportant in low-energy plasmonics. 

\emph{Random-phase approximation.---}To compute the optical response of graphene disks in both TB and Dirac approaches, the first step is to evaluate the non-interacting polarizability~\cite{Thongrattanasiri:2012,Manjavacas:2014}:
\begin{equation}\label{eq:polarizability}
\chi^0(\rv,\rv';\omega) = 2\!\sum_{\nu\nu'}(f_{\nu} - f_{\nu'}) \frac{\boldsymbol{\psi}_{\nu'}^{\dagger}(\rv)\boldsymbol{\psi}_{\nu}(\rv)\boldsymbol{\psi}_{\nu}^{\dagger}(\rv')\boldsymbol{\psi}_{\nu'}(\rv')}{\epsilon_{\nu}-\epsilon_{\nu'}-\hbar(\omega+i\eta)},
\end{equation}
where $f_{\nu}$ denotes Fermi--Dirac equilibrium functions evaluated at energy $\epsilon_{\nu}$, and electron relaxation is included phenomenologically through a finite rate $\eta$.
We give explicit expressions for the Dirac-disk polarizability in the SM.

The random-phase approximation (RPA) is instated by coupling the induced charge density $\rho(\rv)$ to the \emph{total} field via $\chi^0$, leading to a self-consistent integral equation, reading, in operator-notation, as $\boldsymbol{\rho} = e^2\boldsymbol{\chi}^0(\boldsymbol{\phi}^{\mathrm{ext}}+\mathbf{V}\boldsymbol{\rho})$, with $\mathbf{V}$ denoting the Coulomb interaction and $\boldsymbol{\phi}^{\mathrm{ext}}$ an external potential~\cite{GrossoParravicini}. Henceforth, depending on the choice for single-particle states used in constructing $\chi^0$, we distinguish between approaches by the self-evident notation RPA@Dirac and RPA@TB. In the SM we elucidate the technical details for efficiently computing RPA@Dirac via an angular momentum decomposition, and follow the scheme introduced in Ref.~\onlinecite{Thongrattanasiri:2012} for RPA@TB. 
The optical absorption cross-section, i.e.\ the absorbed power relative to the intensity of an incident plane-wave, relates to the induced charge density via $\sim \omega\,\mathrm{Im}[\mathrm{p}(\omega)]$, with $\mathrm{p}(\omega)$ denoting the dipole moment obtained from $\rho(\rv)$.

\emph{Local response.---}For comparison with the quantum approaches described above, we consider also the traditional, classical approach, wherein the induced charge density in graphene is determined from the well-known bulk local-response (LR) conductivity with intra- and interband terms $\sigma_{\text{\textsc{b}}}(\omega) = \sigma_{\mathrm{intra}}(\omega)+\sigma_{\mathrm{inter}}(\omega)$~\cite{Falkovsky:2007}. The interband term induces a redshift~\cite{Wang:2013a} of the dipolar plasmon resonance with decreasing radius, but not to the extent observed in TB-RPA calculations~\cite{Thongrattanasiri:2012}. For the electrostatic disk, the LR problem is solved most elegantly by using a polynomial expansion technique, as explicated by Fetter~\cite{Fetter:1986}, and summarized for completeness in the SM, allowing a semi-analytical solution requiring only a numerical matrix inversion. Applying this technique, we find that the singly radially quantized dipole plasmon resonance, $\omega_{\mathrm{dp}}$, being the resonance of primary relevance in nanoscopic disks, relates to the total LR conductivity $\sigma(\omega)$ via $\omega_{\mathrm{dp}}/\sigma(\omega_{\mathrm{dp}}) = \zeta/2i\varepsilon_0\varepsilon_{\text{\textsc{b}}}R$, with $\varepsilon_{\text{\textsc{b}}}$ denoting the background dielectric constant and $\zeta \approx 1.0977$ accounting for the disk geometry~\cite{Note3}.

\begin{figure}
\hspace*{-.5em}\includegraphics[scale=.93]{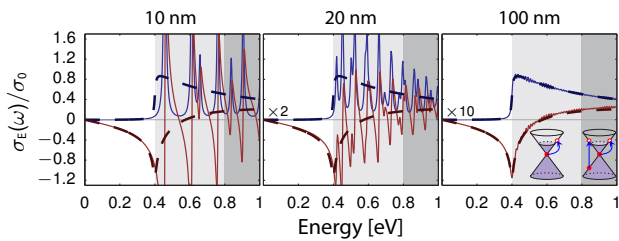} 
\caption{Edge-state conductivity in graphene nanodisks (doped to 0.4 eV) normalized to $\sigma_0 = e^2/4\hbar$ calculated with finite damping $\hbar\eta=6\,\text{meV}$. Disk diameter indicated in titles; note the scaling-factor in the center and right-hand graphs. 
Explicit evaluation of Eq.~\eqref{eq:sigmaedge_full} given in full, and large-radius limit, Eq.~\eqref{eq:sigmaedge_approx}, in bold dashed; real and imaginary parts in blue and red, respectively. 
The region of edge-to-bulk transitions is indicated in gray, while the region of concurrent edge-to-bulk and interband transitions is dark-gray, and illustrated schematically.} \label{fig:edgeconductivity}
\end{figure}

Although the bulk LR conductivity $\sigma_{\text{\textsc{b}}}(\omega)$ is usually derived from a starting point of a continuum of bulk graphene Dirac states, it may as well be derived from the large-radius limit of the finite sample's conductivity using the states $\bm{\psi}_{ln}^{\kappa}$ from Eq.~\eqref{eq:bulkstates}. 
Specifically, in the LR limit, the current response due to an $x$-polarized field is obtained from the conductivity~\cite{GrossoParravicini}:
\begin{equation}\label{eq:conductivity_dipoleapprox}
\sigma(\omega) =\frac{ 2ie^2\omega}{\mathcal{A}} \sum_{\nu\nu'} (f_{\nu}-f_{\nu'})\frac{|\langle \bm{\psi}_{\nu}|x\hspace*{.05em} |\bm{\psi}_{\nu'}\rangle|^2}{\epsilon_{\nu}-\epsilon_{\nu'}-\hbar(\omega+i\eta) }.
\end{equation}
Considering the Dirac ZZ states in Eqs.~\eqref{eq:bulkstates} and~\eqref{eq:edgestates} this gives rise to two distinct terms, one tending asymptotically to $\sigma_{\text{\textsc{b}}}(\omega)$ with increasing radius, originating from bulk-to-bulk transitions $|\langle \bm{\psi}_{ln}^{\kappa}|x\hspace*{.05em}|\bm{\psi}_{l'n'}^{\kappa}\rangle|^2$, and one novel term originating from edge-to-bulk transitions:
\begin{equation}
\sigma_{\text{\textsc{e}}}(\omega) = \frac{4ie^2\omega}{\mathcal{A}}\sum_{\kappa l n \ell}(f_{ln}-f_0)\frac{|\langle \bm{\psi}_{ln}^\kappa|x\hspace*{.05em}|\bm{\phi}_{\ell}^{\kappa}\rangle|^2}{\epsilon_{ln}^2-\hbar^2(\omega+i\eta)^2},
\end{equation}
with $f_0$ denoting the Fermi--Dirac function at zero energy, and $\epsilon_{ln}$ denoting the Dirac ZZ energies corresponding to $\bm{\psi}_{ln}^{\kappa}$. This edge-contribution, physically representing all interactions between occupied zero-energy edge states and unoccupied nonzero-energy bulk states, can be worked out explicitly as a summation over the Bessel function zeros $\beta_{ln}$, see SM for details. In the low-temperature limit, assuming positive $\ef$, the edge-state conductivity becomes:
\begin{equation}\label{eq:sigmaedge_full}
\sigma_{\text{\textsc{e}}}(\omega) = \frac{-16ie^2}{\pi\hbar}\frac{\omega}{\omega_R}
\sum_{\ell=0}^{\ell_{\mathrm{max}}}\sum_n^{\hbar\omega_R\beta_{\ell n}>\ef}
\!\!\!\!
\frac{\ell+1}{\beta_{\ell n}^5\Big[1-\Big(\frac{\omega+i\eta}{\beta_{\ell n}\omega_R}\Big)^2\Big]}.
\end{equation}
Remarkably, the above expression allows a simple asymptotic form in the large-radius limit $R\rightarrow\infty$. Replacing the angular momenta $\ell+1$ by their average value at fixed energy $\ell+1\rightarrow\langle \ell + 1\rangle_{\epsilon} \simeq \xi\epsilon/\hbar\omega_R$, with proportionality constant $\xi = 4/3\pi$ (see SM), introducing the bulk-energy substitution $\beta_{\ell n}\rightarrow \epsilon/\hbar\omega_R$, and transforming the $\ell n$-summations into integrals over $\ef\leq\epsilon<\infty$, we find:
\begin{equation}\label{eq:sigmaedge_approx}
\sigma^{\infty}_{\text{\textsc{e}}}(\omega) =\xi \frac{2e^2}{\pi\hbar}\frac{\vf}{\omega R}
\Bigg[ i\ln \bigg|\frac{\epsilon_{\text{\textsc{f}}}^2-\hbar^2\omega^2}{\epsilon_{\text{\textsc{f}}}^2}\bigg|
+ \pi\theta(\hbar\omega-\epsilon_{\text{\textsc{f}}})\Bigg],
\end{equation}
shown here, for simplicity, in the low-loss limit $\eta\rightarrow 0^+$~\cite{Note4}.
Interestingly, the inclusion of edge states opens a dispersive channel scaling with $\omega_R = \vf/R$, mathematically reminiscent of, but physically distinct from, the scaling phenomenologically introduced in Kreibig damping~\cite{Kreibig:1969} and recently derived from the viewpoint of nonlocal diffusion-dynamics~\cite{Mortensen:2014}.
In addition to Landau damping due to vertical transitions, as included in $\sigma_{\mathrm{inter}}(\omega)$ for $\hbar\omega>2\ef$, edge-to-bulk transitions allow non-vertical transitions at $\hbar\omega>\ef$, with the necessary momentum supplied by the structural truncation with a strength proportional to $1/R$. In Fig.~\ref{fig:edgeconductivity} we consider $\sigma_{\text{\textsc{e}}}(\omega)$ and compare with $\sigma_{\text{\textsc{e}}}^\infty(\omega)$ for three disk diameters. At smaller diameters $\sigma_{\text{\textsc{e}}}(\omega)$ and $\sigma_{\text{\textsc{e}}}^\infty(\omega)$ differ substantially in the region $\hbar\omega>\ef$ with $\sigma_{\text{\textsc{e}}}$ exhibiting peaks at discrete transitional energies $\hbar\omega \simeq \epsilon_{\ell n}$; as the diameter is increased, and the energy difference between distinct transitional energies decrease accordingly, $\sigma_{\text{\textsc{e}}}$ approaches $\sigma_{\text{\textsc{e}}}^\infty$ asymptotically, as anticipated. We note that a generally good agreement is apparent, even for small disks, when $\hbar\omega<\ef$. The importance of the edge-state conductivity vis-\'{a}-vis the bulk-conductivity diminishes with increasing diameter due to the $1/R$ scaling of $\sigma_{\text{\textsc{e}}}(\omega)$. Nonetheless, even at large disk diameters, e.g.\ 20~nm, the maximal edge-state conductivity is still on the order of $\sim\!\!0.4\sigma_0$, while the magnitude of vertical interband transitions roughly amounts to $\sigma_0$.

\emph{Hydrodynamic response.---} The non-interacting polarizability, the key constituent of the RPA, as considered in Eq.~\eqref{eq:polarizability}, accounts not only for the discretized and individual nature of the allowable states, but also for the nonlocal nature of the electromagnetic response, manifest in the finitude of $\chi^0(\rv,\rv';\omega)$ for $\rv\neq\rv'$. An approximate accounting of nonlocal response can be facilitated by a hydrodynamic model~\cite{Wang:2013a}:
\begin{equation}
\bigg(1+\frac{\beta^2}{\omega^2}\nabla_{\scriptscriptstyle\parallel}^2\bigg)\,\rho(\rv) = \frac{i\sigma(\omega)}{\omega}\nabla_{\scriptscriptstyle\parallel}^2\phi(\rv),
\end{equation}
with $\nabla_{\scriptscriptstyle\parallel}^2$ being the two-dimensional Laplacian, and with plasma velocity denoted by $\beta^2=\tfrac{3}{4}\vf^2$, see SM. For brevity, we will denote hydrodynamic calculations with a backbone conductivity $\sigma(\omega)$ by $[\sigma]^{\text{\textsc{h}}}(\omega)$. The primary effect of the hydrodynamic model is to introduce a blueshift, which, in $[\sigma_{\text{\textsc{b}}}]^{\text{\textsc{h}}}(\omega)$, approximately amounts to a shift $\delta\omega_{\mathrm{dp}}\simeq 1.27\omega_R^2/\omega_{\mathrm{dp}}$. Predictions of the hydrodynamic model at the level $[\sigma_{\text{\textsc{b}}}]^{\text{\textsc{h}}}(\omega)$ agree excellently with predictions of RPA@Dirac IM as we show in the SM. This underscores the accuracy of a hydrodynamic description, since RPA@Dirac IM neglects the existence of edge states, and thus, at large radii, is modified primarily by nonlocal effects.

\begin{figure}[!htb]
\hspace*{-.3em}\includegraphics[scale=.93]{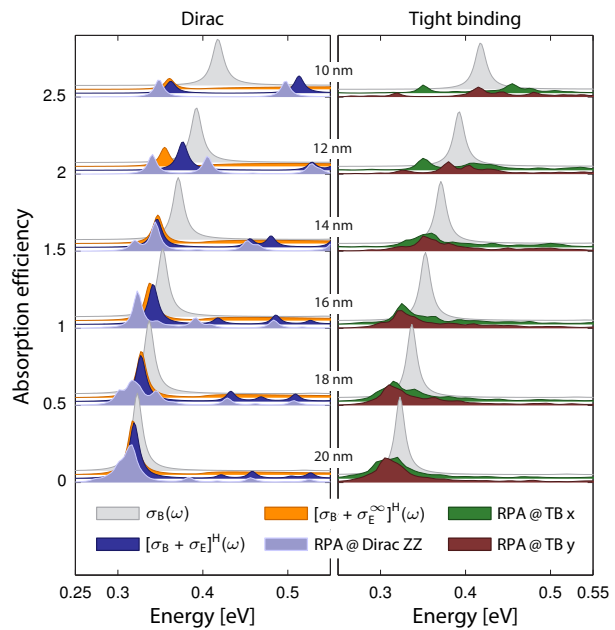} 
\caption{Absorption cross-sectional efficiency of graphene nanodisks calculated via LR, hydrodynamics with bulk and edge-state conductivities, RPA@Dirac ZZ, and RPA@TB (for $x$- and $y$-polarized light) for disks of increasing diameter. The sample is considered doped to $\ef = 0.4\ \mathrm{eV}$, with electron relaxation-rate $\hbar\eta=6\ \mathrm{meV}$, and at a temperature $T=300\ \mathrm{K}$. Spectra for different diameters are offset by 0.5, while individual spectra at identical diameters are offset by 0.025. Spectra for intermediate diameters available in SM.}\label{fig:absorption}
\end{figure}

\emph{Results and discussion.---} Figure \ref{fig:absorption} depicts the absorption cross-sectional efficiency, i.e.\ cross-section normalized to disk area, of graphene nanodisks for different diameters, contrasting results obtained by LR with and without hydrodynamic and edge-state conductivity, RPA@TB, and RPA@Dirac ZZ. A key feature of RPA@TB, not captured by any of the continuum models, is a polarization-dependence of the optical response due to the discrete nature of the description. For smaller disks, only few EHPs contribute, leading to a strong polarization-dependence. For larger disks, as the number of contributing EHPs increase, and the collective nature of the plasmon emerges, this dependence diminishes rapidly.

The primary feature of both RPA@TB and RPA@Dirac ZZ for disk-diameters larger than approximately 14 nm, is the emergence of a broad dominant plasmonic resonance redshifted with respect to the LR bulk predictions. Comparison with $[\sigma_{\tsc{b}}+\sigma_{\tsc{e}}]^{\tsc{h}}(\omega)$ and $[\sigma_{\tsc{b}}+\sigma_{\tsc{e}}^{\infty}]^{\tsc{h}}(\omega)$ agrees qualitatively. A similar redshift is reproduced, $\appropto\!\omega_R^2/\omega_{\mathrm{dp}}\appropto\! 1/R$ (see SM), but slightly underestimated in magnitude due to the assumption of a constant total field, inherent to the dipole-approximation in Eq.~\eqref{eq:conductivity_dipoleapprox}, contrasting the actual electric field distribution of the plasmon, which is significantly concentrated near the edge~\cite{Wang:2011}. 
Furthermore, the dipole resonance in RPA@TB is damped and broadened to a larger degree than both RPA@Dirac ZZ and $[\sigma_{\tsc{b}}+\sigma_{\tsc{e}}]^{\tsc{h}}(\omega)$ as a result of the explicit breaking of azimuthal and valley symmetry in the discrete treatment, permitting additional dipole-allowed transitions.

In conclusion, the redshift observed between predictions of RPA@TB and bulk LR calculations arises from the competing effects of edge-conductivity and nonlocal response, with the former prevailing, shifting the dipole resonance $\omega_{\mathrm{dp}}$ to the red and blue, respectively, with a strength $\appropto\! \omega_R^2/\omega_{\mathrm{dp}}$ in both cases. The simultaneous accounting of both effects is thus of paramount importance in semi-classically reproducing the key plasmonic features of full RPA@TB predictions, with significant corrections even at relatively large diameters $\sim\! 20\ \text{nm}$. The identification of this additional dispersion channel via edge-states extends and enriches the plethora of already identified nonclassical-corrections for plasmonics at the nanoscale. The finding of a simple analytical term accounting for optical response of edge states, Eq.~\eqref{eq:sigmaedge_approx}, offers the promise of extendability to other geometries in graphene nanostructures, e.g.\ ribbons or triangles, as well as to a wider class of systems supporting edge or surface states in general, e.g.\ in topological insulators~\cite{Hasan:2010}, MoS$_2$ nanotriangles~\cite{Bollinger:2003}, nanostructures with Ag(111) facets~\cite{Hsieh:1985}, or even in systems with highly localized defect states. 

\vskip .3em
\begin{acknowledgments}
\emph{Acknowledgments.---}We thank Wei Yan for many fruitful discussions, helpful suggestions, and continued interest.
The Center for Nanostructured Graphene is sponsored by the Danish National Research Foundation, Project DNRF58.
This work was also supported by the Danish Council for Independent Research--Natural Sciences, Project 1323-00087.
\end{acknowledgments}


\end{document}